\documentclass[preprint,3p,12pt]{elsarticle}

\usepackage{amssymb}
\usepackage{amsmath}
\usepackage{mathtools}
\usepackage{amsfonts}
\usepackage{amsbsy}
\usepackage{color}
\usepackage{float}
\usepackage{graphicx}
\usepackage{caption}
\usepackage{subcaption}
\usepackage{lineno}
\usepackage{comment}
\usepackage{algorithm}
\usepackage{algorithmic}
\usepackage{tcolorbox}
	
\newcommand{\bs}{\boldsymbol}

\newcommand{\bm}{\mathbf}
\newcommand{\mrm}{\mathrm}
\newcommand{\domain}[1][]{
	\ensuremath{\mathcal{H}^{#1}}}

\newcommand{\state}{\underline}
\newcommand{\base}[1][\boldsymbol{\xi}]{\langle#1\rangle}
\newcommand{\dV}[1][\boldsymbol{\xi}]{\,\mathrm{d}V_{#1}}

\newcommand{\unity}[1][2]{
	\ensuremath{\mathbb{I}^{(#1)}}
}

\begin{document}

\begin{frontmatter}

\title{A Unified Framework of Bond-Associated Peridynamic Material Correspondence Models}

\author[ucb]{Xuan Hu}
\ead{xuan_hu@berkeley.edu}

\author[uky]{Hailong Chen}
\ead{hailong.chen@uky.edu}

\author[nyu1]{Yichi Zhang}
\ead{yz7996@nyu.edu}

\author[nyu2]{Zening Wang}
\ead{zw3079@nyu.edu}

\address[ucb]{Department of Civil and Environmental Engineering, University of California, Berkeley, CA 94720, USA}

\address[uky]{Department of Mechanical and Aerospace Engineering, University of Kentucky, Lexington, KY 40506, USA}

\address[nyu1]{College of Arts \& Science, New York University, New York, NY 10003, USA}
\address[nyu2]{Tandon School of Engineering, New York University, Brooklyn, NY 11201, USA}

\begin{abstract}
This paper presents a unified framework for bond-associated peridynamic material correspondence models that were proposed to inherently address the issue of material instability or existence of zero-energy modes in the conventional correspondence formulation. The conventional formulation is well-known for having the issue of material instability due to the non-unique mapping between bond force density state and nonlocal deformation gradient. Several bond-associated models that employ bond-level deformation gradients address this issue in a very effectively and inherent manner. Although different approaches were taken to formulate bond-level deformation gradient so the bond-associated quantities can be captured more accurately, a detailed study finds a unified systematic framework exists for these models. It is the purpose of this paper to consolidate these approaches by providing a unified and systematic framework for bond-associated peridynamic correspondence models. Based on all the bond-associated deformation gradients proposed in the literature, a unified bond-associated deformation gradient is formulated. Assuming energy equivalence with the local continuum mechanics theory, the unified bond force density state is derived using the Fr\'echet derivative. Additionally, the properties of the formulated unified framework including linear momentum balance, angular momentum balance, and objectivity are thoroughly examined. This work serves as a valuable reference for the further development and application of bond-associated correspondence formulations in peridynamics. 
\end{abstract}



\end{frontmatter}

\section{Introduction}
Peridynamics is a nonlocal continuum mechanics theory that addresses the limitations of the classical local theory in dealing with spatial discontinuities and accounting for length scale effects~\cite{silling2000bond, silling2007state,bobaru2012horizon,bobaru2016handbook,chen2020higher,chan2023higher}. The development of peridynamics began with the seminal work by Silling~\cite{silling2000bond} on reformulation of elasticity theory for discontinuities and long-range forces, where pairwise bond-based interactions within finite distance called horizon are formulated. In this bond-based formulation, the force density of a bond depends only on its stretch. While it is effective in capturing fracture phenomena, the bond-based formulation is limited in describing general material behaviors such as arbitrary Poisson ratio and nonlinear constitutive relationship, due to the usage of a central potential that is totally independent of all other local conditions~\cite{silling2007state}. To overcome this limitation, the state-based formulation that rewrites the material-dependent part of the peridynamic model using the concept of state was introduced~\cite{silling2007state}. More importantly, the material correspondence formulation, a subset of the state-based generalization, bridges the gap between peridynamics and the classical continuum mechanics theory by allowing direct incorporation of continuum material models into peridynamics. This is achieved by introducing nonlocal deformation gradient and stress tensors in a manner equivalent to the classical continuum mechanics theory but within a nonlocal framework.

However, the material correspondence formulation is not without challenges. One well-known issue of the formulation is the existence of material instability or zero-energy modes manifested in the form of oscillation in the displacement field. These modes arise when certain deformation states do not contribute to the strain energy, leading to non-physical solutions and numerical instabilities. Among existing strategies proposed in the literature to address this issue, the bond-associated formulations are the most effective and provide more accurate accounting of bond-level quantities such as deformation gradient and stresses. In all the bond-associated formulations, the nonlocal deformation gradient is constructed for bond instead of material point. This is the main difference between the bond-associated formulation with the conventional formulation proposed by Silling et al. The bond-associated deformation gradients are more suitable for and accurate in capturing the deformation of each individual bond. As a result, the map from the bond deformation to the bond force state is injective and material instability or zero-energy modes are inherently eliminated. The bond-associated deformation gradient and corresponding material correspondence model was first introduced by Chen~\cite{chen2018bond,chen2019stability,chan2021wave,chan2022reformulation}. In this formulation, a bond-associated deformation gradient is constructed for each bond within the horizon based on a subset of the horizon. In general, each bond has its own unique subset, and when the subset takes the whole horizon, the conventional material correspondence model is recovered. Chowdhaury et al.~\cite{chowdhury2019modified} proposed to partition the horizon into sub-horizons and construction of the nonlocal deformation gradient is limited to each sub-horizon. Within each sub-horizon, the same nonlocal deformation gradient is used for all bonds within that sub-horizon. Although these two formulations differ slightly, the core idea is to use a subset of the horizon that includes the target bond to characterize the bond deformation and compute the bond-associated deformation gradient. The formulation proposed by Chowdhaury can be considered as a special case of the formulation proposed by Chen. Breitzman and Dayal~\cite{breitzman2018bond} proposed a bond-level deformation gradient by firstly removing the contribution due to the uniform deformation assumption within the horizon using deformation mapping and replacing it with the actual bond deformation. Hou and Zhang~\cite{hou2024bond} developed the so-called bond-augmented deformation gradient, where a penalty term related to a given bond and its deformed state is introduced during the minimization of the least squares error to formulate the nonlocal deformation gradient~\cite{chen2018bond}. In these two work, the nonlocal deformation gradients are still constructed based on the whole horizon, but they are modified to be bond-specific either through replacement~\cite{breitzman2018bond} or penalization~\cite{hou2024bond}. Bond-associated deformation gradients can also be obtained using non-spherical influence function. Chen et al.~\cite{chen2023influence, chen2024generalized} proposed a family of non-spherical influence function and developed the corresponding material correspondence model to improve the accuracy of bond-level quantities such as deformation gradient and stress. Unlike the conventional formulation, where the influence function is spherical and depends only on the bond length, the proposed non-spherical influence functions take into account both the bond length and the bond relative angle (with respect to a target bond). All these bond-associated correspondence models have achieved great success in inherently eliminating the material instability in the conventional formulation.

Although different approaches were proposed to develop bond-level deformation gradients to eliminate material instability or zero-energy modes, a detailed examination of these bond-associated formulations finds that a unified framework is shared among them. It is the goal of this paper to present this unified framework for bond-associated peridynamics material correspondence models and examine the physical properties of the framework. This framework will not only address the zero-energy modes issue but also enhance the model’s capability to accurately represent the deformation state at the bond level, thus paving the way for more robust and flexible peridynamics simulations. The rest of the paper is organized as follows: Section 2 introduces the fundamental definitions and notations for the state-based formulation. Section 3 presents a generalized formula for bond-associated deformation gradient, which is then tailored to various bond-associated formulae proposed in the literature. Section 4 derives the generalized force density state using the equivalency of strain energy density between classical continuum mechanics theory and peridynamics. Different forms of the bond force density state from the bond-associated formulations are recovered from the generalized formulation. Section 5 outlines the nonlocal equation of motion of bond-associated material correspondence formulation. This section also includes the proofs of linear momentum balance, angular momentum balance and objectivity. Section 6 summarizes the study and highlights the developed unified framework.

\section{Definitions and Notations}
In peridynamics, the geometric domain of interest in referential configuration $\mathcal{B}$ is modeled as an assembly of material points with volume. For a material point $\bm X$, it interacts with its neighboring material points located within a Euclidean distance $\delta$, which is known as \textit{horizon}, through nonlocal interactions as bond forces. The point within the horizon is called \textit{neighbor} and the collection of all neighbors is referred to as \textit{neighborhood}, denoted as $\domain$. The \textit{relative position in reference configuration} $\mathcal{B}$ between a material point $\bm X$ and its neighbor $\bm X'$ is a bond as $\bs\xi = \bm X' - \bm X$. Let $\bm y(\bm X, t)$ represents a new position of the material point $\bm X$ in the current configuration $\mathcal{B}_t$, with time $t\geq 0$. The \textit{relative position in current configuration} $\mathcal{B}_t$ between two neighboring points is $\bs\zeta = \bm y'(\bm X', t) - \bm y(\bm X, t)$.

The development of the peridynamic material correspondence formulation introduces the concept of state~\cite{silling2007state}. A \textit{state} of order $m$ is a function $\state{\bm A}\base[\bullet]:\domain\rightarrow \mathcal{L}_m$, which maps a vector in the neighborhood $\domain$ to the tensor space  $\mathcal{L}_m$ of order $m$. For instance, if $m=0$, a bond is mapped to a scalar space $\mathcal{L}_0$, and this is referred to as a scalar state. Scalar states are usually written in lowercase, non-boldface with an underscore, such as $\state{\omega}\base, \state{\mrm{w}}\base$. If $m=1$, a bond is mapped to a vector space $\mathcal{L}_1$, and the state is called vector state. vector states and other states of order $m\geq 1$ are conventionally written in uppercase boldface with an underscore, such as $\state{\bm y}\base$ or $\state{\bm A}\base$. According to this definition, a state can be readily identified that maps the initial bond vector $\bs\xi$ to the current bond vector $\bs\zeta$. This is termed the \textit{deformed state} and denoted by $\state{\bm y}\base$. For a bond connecting a material point $\bm X$ and one of its neighbors at any time $t$, the deformed state is $\state{\bm y}[\bm X, t]\base = \bm y'(\bm X', t) - \bm y(\bm X, t) = \bs\zeta$. For clarity, square brackets are introduced to indicate the spatial or temporal information on which a state depends, e.g., $[\bm X, t]$, while parentheses are adopted to denote all other quantities that a state depends on, e.g., $(\bs\xi, \bs\xi')$. In addition, the standard convention in classical continuum mechanics theory is followed in this paper. Variables with uppercase subscripts, such as $X_I$, refer to the components defined in reference configuration $\mathcal{B}$ while the ones with lowercase subscripts, such as $y_j$ denote the components in current configuration $\mathcal{B}_t$. Einstein summation notation is also employed here to facilitate the representation of complex tensor operations. Boldface letters indicate vectors or tensors.

\section{Unified Bond-Associated Nonlocal Deformation Gradient}
To start the formulation of a unified framework of peridynamic bond-associated correspondence models, the generalized formula for computing the bond-associated nonlocal deformation gradient is developed in this section. For a given bond $\bs\xi'$, the generalized bond-associated deformation gradient has the following expression as
\begin{tcolorbox}
    \begin{align}
        \bm F_{\bs \xi'} = \left[\int_{\domain} \state\omega\base \state{\bm y}\base \otimes \bs\xi\dV\right]\bm K_{\bs \xi'}^{-1}\state{\bm A}\langle\bs\xi'\rangle + \state{\bm B}\langle\bs\xi'\rangle
        \label{eq:F_xi_general}
    \end{align}
    with
    \begin{align}
        \bm K_{\bs \xi'} = \int_{\domain} \state\omega \bs\xi \otimes \bs\xi\dV,
        \label{eq:K_shape}
    \end{align}
\end{tcolorbox}
\noindent
where the subscript $\bs\xi'$ indicates that the quantities are associated with the bond $\bs\xi'$, $\state{\omega}\base$ is a state-valued influence function that satisfies $\int_{\domain} \state\omega\base \dV = 1$, $\state{\bm y}\base$ denotes the deformed state of bond vector $\bs\xi$ within the horizon $\domain$, and $\otimes$ indicates tensor product. This generalized formula can be adapted into various forms proposed in the literature by selecting $\state{\bm A}\langle\bs\xi'\rangle$ and $\state{\bm B}\langle\bs\xi'\rangle$.

\begin{itemize}
    \item \textbf{Conventional model}
\end{itemize}

In the conventional correspondence model proposed by Silling et al.~\cite{silling2007state}, the nonlocal deformation gradient can be obtained from Eq.~\eqref{eq:F_xi_general} by adopting a spherical influence function that depends solely on bond length $|\bs\xi|$ and taking state $\state{\bm A}\base[\bs\xi']$ as the identity tensor and the state $\state{\bm B}\base[\bs\xi']$ as the null tensor, i.e.,
\begin{align}
    \state\omega \langle\bs\xi\rangle \Rightarrow \state\omega(|\bs\xi|)\langle\bs\xi\rangle, \quad
    \state{\bm A}\langle\bs\xi'\rangle\Rightarrow \unity[2], \quad
    \state{\bm B}\langle\bs\xi'\rangle\Rightarrow \bm 0.
    \label{eq:Cond_Conv}
\end{align}
Therefore, the conventional nonlocal deformation gradient has the following expression
\begin{align}
    \bm F_{\bs \xi'} = \left[\int_{\domain}\state\omega(|\bs\xi|)\base\state{\bm y} \otimes \bs\xi \dV\right]\bm K_{\bs\xi'}^{-1}
    \label{eq:dg_Conv}
\end{align}
with $\bm K_{\bs\xi'}$ given in Eq.~\eqref{eq:K_shape}.

It should be noted that even though the nonlocal deformation gradient $\bm F_{\bs\xi'}$ is supposed to be associated with bond $\bs\xi'$, it remains the same for all bonds within the same horizon. As has been pointed out in the literature~\cite{chen2018bond,chowdhury2019modified}, the mapping between the conventional nonlocal deformation gradients and the deformation states is non-injective. This implies that one nonlocal deformation gradient computed by Eq.~\eqref{eq:dg_Conv} can correspond to more than one deformation states within a given horizon. As a result, there exists non-zero deformation states that rise zero energy to the system, i.e., existence of zero-energy modes in the solution.

\begin{itemize}
    \item \textbf{Sub-horizon-based models}
\end{itemize}

To address the issue of the zero-energy modes in the conventional model, Chen et al.~\cite{chen2018bond,chen2019stability,chan2021wave,chan2022reformulation} and Chowdhaury et al.~\cite{chowdhury2019modified} proposed the sub-horizon-based stabilization approach. Despite of the slight difference, the key idea behind their models is to use a subset of horizon that includes the bond $\bs\xi'$ to characterize the deformation of bond $\bs\xi'$ and compute the bond-associated deformation gradient. In comparison with conventional model, the influence function, state $\state{\bm A}\base[\bs\xi']$ and state $\state{\bm B}\base[\bs\xi']$ remain unchanged. The only modification is that the integration domain is reduced from the full horizon $\domain$ to a sub-horizon $h_{\bs\xi'}$, i.e.,
\begin{align}
    \state\omega \langle\bs\xi\rangle \Rightarrow \state\omega(|\bs\xi|)\langle\bs\xi\rangle,\quad
    \domain\Rightarrow h_{\bs\xi'},\quad
    \state{\bm A}\langle\bs\xi'\rangle\Rightarrow \unity[2],\quad
    \state{\bm B}\langle\bs\xi'\rangle\Rightarrow \bm 0.
    \label{eq:Cond_Sub}
\end{align}

As a result, the bond-associated deformation gradient for the sub-horizon-based models has the following expression
\begin{align}
    \bm F_{\bs \xi'} =& \left[\int_{h_{\bs\xi'}}\state\omega(\bs\xi)\state{\bm y} \otimes \bs\xi\dV\right]\bm K_{\bs\xi'}^{-1}
\end{align}
with
\begin{align}
    \bm K_{\bs\xi'} = \int_{h_{\bs\xi'}} \state\omega(\bs\xi) \bs\xi \otimes \bs\xi\dV
\end{align}

\begin{itemize}
    \item \textbf{Projection-based model}
\end{itemize}

Breitzman and Dayal~\cite{breitzman2018bond} proposed a bond-level deformation gradient by first eliminating the contribution of the uniform nonlocal deformation on bond $\bs\xi'$ through the use of a projection tensor and then replacing it with the actual deformation. In this approach, the influence function remains a spherical state-valued function, while the state $\state{\bm A}\base[\bs\xi']$ and state $\state{\bm B}\base[\bs\xi']$ are replaced by the projection tensor and actual deformation respectively, i.e., 
\begin{align}
    \state\omega \langle\bs\xi\rangle\Rightarrow \state\omega(|\bs\xi|)\langle\bs\xi\rangle,\quad
    \state{\bm A}\langle\bs\xi'\rangle\Rightarrow \unity[2] - \frac{\bs\xi'\otimes\bs\xi'}{|\bs\xi'|^2},\quad
    \state{\bm B}\langle\bs\xi'\rangle\Rightarrow \frac{\state{\bm y}' \otimes \bs\xi'}{|\bs\xi'|^2}.
    \label{eq:Cond_Proj}
\end{align}
Hence, the expression for this bond-associated deformation gradient is
\begin{align}
    &\bm F_{\bs \xi'} = \left[\int_{\domain}\state\omega(\bs\xi)\state{\bm y} \otimes\bs\xi\dV\right]\bm K_{\bs\xi'}^{-1}\left(\unity[2] - \frac{\bs\xi'\otimes\bs\xi'}{|\bs\xi'|^2} \right) + \frac{\state{\bm y}' \otimes\bs\xi'}{|\bs\xi'|^2}
\end{align}
with $\bm K_{\bs\xi'}$ given in Eq.~\eqref{eq:K_shape}.

This model not only eliminate the non-physical deformations such as interpenetration and material instability issues, but also accurately represent both the average deformation within one neighborhood and the stretch of specific bond~\cite{breitzman2018bond}.

\begin{itemize}
    \item \textbf{Lagrangian-multiplier-based model}
\end{itemize}

Hou and Zhang~\cite{hou2024bond} constructed a so-called bond-augmented deformation gradient in their work. Similar to Lagrangian multiplier method, a penalty term related to $\bs\xi'$ and its deformed state $\state{\bm Y}\base[\bs\xi']$ is introduced during the minimization of the least squares error. In this approach, the state $\state{\bm A}\base[\bs\xi']$ and $\state{\bm B}\base[\bs\xi']$, as well as the horizon, becomes
\begin{align}
    \state\omega \langle\bs\xi\rangle \Rightarrow \state\omega(|\bs\xi|)\langle\bs\xi\rangle,\quad
    \domain\Rightarrow \domain\backslash \bs\xi',\quad
    \state{\bm A}\base[\bs\xi']\Rightarrow \unity[2],\quad
    \state{\bm B}\base[\bs\xi']\Rightarrow \lambda\int_{\domain}\delta\left[\bs\xi-\bs\xi' \right] \state\omega(\bs\xi) \state{\bm y} \otimes\bs\xi\cdot \bm K_{\bs\xi'}^{-1}\dV,
    \label{eq:Cond_Lag}
\end{align}
where $\lambda$ is the penalty factor and $\delta[\cdot]$ is the delta function where $\delta[\bm x]=1$ when $\bm x = 0$ and $\delta[\bm x]=0$ otherwise. As a result, the bond-associated deformation gradient $\bm F_{\bs\xi'}$ and shape tensor $\bm K_{\bs\xi'}$ are transformed into the following expressions.
\begin{align}
    &\bm F_{\bs \xi'} = \left[\int_{\domain\backslash \bs\xi'} \state\omega(\bs\xi)\state{\bm y} \otimes \bs\xi\dV\right]\bm K_{\bs\xi'}^{-1} + \lambda\int_{\domain}\delta\left[\bs\xi-\bs\xi' \right]\state\omega(\bs\xi) \state{\bm y} \otimes \bs\xi\cdot\bm K_{\bs\xi'}^{-1}\dV
    \label{eq:def_F_Lag}
\end{align}
and
\begin{align}
    \bm K_{\bs\xi'} = \int_{\domain\backslash \bs\xi'} \state\omega(\bs\xi) \bs\xi \otimes \bs\xi\dV + \lambda\int_{\domain}\delta\left[\bs\xi-\bs\xi' \right]\state\omega(\bs\xi) \bs\xi \otimes \bs\xi\dV.
    \label{eq:def_K_Lag}
\end{align}

It is evident that the difference between $\bm F_{\bs\xi'}\bs\xi'$ and $\state{\bm y}\base[\bs\xi']$ decreases as the penalty factor $\lambda$ increases. However, an extremely large $\lambda$ may result in a singular shape tensor leading to numerical error. Therefore, a reasonably large penalty factor should be chosen in practice.

In the model proposed by Hou and Zhang~\cite{hou2024bond}, the penalty term together with the penalty factor is explicitly expressed in the Eqs.~\eqref{eq:def_F_Lag} and \eqref{eq:def_K_Lag}. However, it is more convenient to incorporate the penalty term into the influence function and reformulate the equations given above. Let the influence function be
\begin{align}
    \state{\omega}(\bs\xi, \bs\xi')\base = 
    \begin{dcases}
        \lambda \omega_0,\quad &\bs\xi = \bs\xi'\\
        \omega_0,\quad &\bs\xi \neq \bs\xi'
    \end{dcases},
    \label{eq:new_weight}
\end{align}
where $\omega_0$ is a constant and the condition
\begin{align}
    \int_{\domain}\state{\omega}(\bs\xi, \bs\xi')\base \dV = 1 
\end{align}
should be still satisfied.

In this sense, the deformation gradient $\bm F_{\bs\xi'}$ associated to bond $\bs\xi'$ can be alternatively expressed as
\begin{align}
    \state\omega\base\Rightarrow \state\omega(\bs\xi, \bs\xi')\base,\quad
    \state{\bm A} \base[\bs\xi']\Rightarrow \unity[2],\quad
    \state{\bm B}\base[\bs\xi']\Rightarrow \bm 0.
    \label{eq:Cond_Lag2}
\end{align}

Finally, the nonlocal deformation gradient can be rewritten as
\begin{align}
    \bm F_{\bs \xi'} = \left[\int_{\domain} \state\omega(\bs\xi,\bs\xi')\base\state{\bm y} \otimes \bs\xi\dV\right]\bm K_{\bs\xi'}^{-1}
    \label{eq:def_F_Lag2}
\end{align}
with
\begin{align}
    \bm K_{\bs\xi'} = \int_{\domain} \state\omega(\bs\xi, \bs\xi')\base \bs\xi \otimes \bs\xi\dV.
    \label{eq:def_K_Lag2}
\end{align}

\begin{itemize}
    \item \textbf{Non-spherical-influence-function-based model}
\end{itemize}

Chen et al.~\cite{chen2023influence,chen2024generalized} proposed a bond-associated deformation gradient by using non-spherical influence function. In contrast to the spherical influence function that depends only on the bond length, the non-spherical influence function depends on both the bond length and relative angle between the bond of interest $\bs\xi'$ and any other bonds within the horizon. This type of bond-associated model can be obtained from Eq.~\eqref{eq:F_xi_general} by using non-spherical influence functions and setting $\state{\bm A}\base[\bs\xi']$ and $\state{\bm B}\base[\bs\xi']$ as $\unity$ and $\bm 0$ respectively, i.e.,
\begin{align}
    \state\omega\base\Rightarrow \state\omega(\bs\xi, \bs\xi')\base,\quad
    \state{\bm A} \base[\bs\xi']\Rightarrow \unity[2],\quad
    \state{\bm B}\base[\bs\xi']\Rightarrow \bm 0,
    \label{eq:Cond_Nonsph}
\end{align}
The non-spherical influence function proposed by Chen et al.~\cite{chen2023influence, chen2024generalized} has the following form as
\begin{align}
    \state\omega(\bs\xi, \bs\xi')\base = \exp\left(-n_1\frac{||\bs\xi| - |\bs\xi'||}{\delta}\right)\left(\frac{1}{2} + \frac{1}{2}\cos(\widehat{\bs\xi\bs\xi'})\right)^{n_2}
    \label{eq:non-spherical}
\end{align}
where $\exp(\cdot)$ is the exponential function, $n_1, n_2 > 0$ are controlling parameters that can control shape of the influence function over the horizon $\domain$, $\widehat{\cdot}$ indicates angle between two bonds.

The bond-associated deformation gradient using the non-spherical influence function shares the same expressions as the Lagrangian-multiplier-based model, i.e., Eqs.~\eqref{eq:def_F_Lag2} and \eqref{eq:def_K_Lag2}, but with the influence function given in Eq. \eqref{eq:non-spherical}. Note that the non-spherical influence function should also satisfy $\int_{\domain} \state\omega(\bs\xi, \bs\xi')\base\dV = 1$. Even though the Lagrangian-multiplier-based model and the non-spherical-influence-function-based model originate from different perspectives, they share one common feature: the influence function assigns varying weights to different bonds within the horizon even for bonds of the same length. In the Lagrangian-multiplier-based model, a higher weight for the bond of interest is explicitly provided by the singular penalty factor whereas in the non-spherical-influence-function-based model, this is achieved through a smooth continuous function.

\section{Unified Force Density State}
The unified force density state corresponding to the unified bond-associated deformation gradient can be derived following the same procedure as outlined by Silling et al.~\cite{silling2007state}. Let's define the nonlocal strain energy density for bond-associated models as:
\begin{equation}
    \mathcal{W} = \int_{\domain}\state{\mrm{w}}\base[\bs\xi'] \bm P_{\bs\xi'}:\bm F_{\bs\xi'}\dV[\bs\xi'],
\end{equation}
\noindent
where $\state{\mrm{w}}\base[\bs\xi']$ is a scalar state-valued weight function that satisfies $\int_{\domain}\state{\mrm{w}}\base[\bs\xi']\dV[\bs\xi']=1$; $\bm F_{\bs\xi'}$ denotes the deformation gradient associated with bond $\bs\xi'$, and $\bm P_{\bs\xi'}$ is the first Piola–Kirchhoff stress (PK1 stress) corresponding to $\bm F_{\bs\xi'}$.

Assuming $\bm F_{\bs\xi'}$ is differentiable, the Fr\'echet derivative of $\bm F_{\bs\xi'}$ can be obtained using the following equation:
\begin{align}
    \bm F_{\bs\xi'}(\state{\bm y}+\Delta\state{\bm y}) =& \left[\int_{\domain}\state\omega\cdot(\state{\bm y} + \Delta\state{\bm y})\otimes \bs\xi\dV\cdot\right]\bm K_{\bs\xi'}^{-1}\state{\bm A}\base[\bs\xi'] + \state{\bm B}\base[\bs\xi'] + \Big(\delta[\bs\xi-\bs\xi']\nabla_{\bm y}\state{\bm B}\base[\bs\xi']\Big) \bullet \Delta\state{\bm y}\nonumber\\
    =& \left[\int_{\domain}\state\omega\cdot\state{\bm y}\otimes \bs\xi\dV\right]\bm K_{\bs\xi'}^{-1}\state{\bm A}\base[\bs\xi'] + \state{\bm B}\base[\bs\xi'] +\nonumber\\
    &\left[\int_{\domain}\state\omega\cdot\Delta\state{\bm y}\otimes \bs\xi\dV\right]\bm K_{\bs\xi'}^{-1}\state{\bm A}\base[\bs\xi']+ \Big(\delta[\bs\xi-\bs\xi']\nabla_{\bm y}\state{\bm B}\base[\bs\xi']\Big) \bullet\Delta\state{\bm y},
    \label{eq:F_y_dy}
\end{align}
where $\bullet$ represents the \textit{dot product} of two states. In state-based peridynamics~\cite{silling2007state}, the dot product of two states $\state{\bm A}\base$ and $\state{\bm B}\base$ of the same order is defined as
\begin{align}
    \state{\bm A}\bullet\state{\bm B} := \int_{\domain} \state{\bm A}\base\cdot \state{\bm B}\base\dV.
\end{align}
From the Fr\'echet derivative provided in Eq.~\eqref{eq:F_y_dy} and by letting $\bm F_{\bs\xi'}(\state{\bm y} + \Delta\state{\bm y}) = \bm F_{\bs\xi'} + \Delta\bm F_{\bs\xi'}$, the increment of deformation gradient $\Delta F_{\bs\xi', iJ}$ can be expressed using index notation as
\begin{align}
    \Delta F_{\bs\xi', iJ} =& \int_{\domain}\state\omega\delta_{il}\xi_Q K^{-1}_{\bs\xi', QR}\state{A}_{RJ}\Delta y_l\dV + \Big(\delta[\bs\xi-\bs\xi'] \nabla_{y_l} \state{B}_{iJ}\Big)\bullet\Delta \state{y}_l\nonumber\\
    =& \left[\state\omega\delta_{il}\xi_QK^{-1}_{\bs\xi', QR}\state{A}_{RJ} + \delta[\bs\xi-\bs\xi'] \nabla_{y_l} \state{B}_{iJ}\right]\bullet\Delta \state{y}_l.
\end{align}

Therefore, the gradient of deformation gradient regarding $\state{y}_l$ is
\begin{align}
    \nabla_{\state{y}_l} F_{\bs\xi', iJ} = \state\omega\delta_{il}\xi_Q K^{-1}_{\bs\xi', QR}\state{A}_{RJ} + \delta[\bs\xi-\bs\xi'] \nabla_{y_l} \state{B}_{iJ}.
\end{align}

It should be noted that this gradient is defined within the framework of state dot product. Similarly, the increment of the nonlocal strain energy density $\Delta\mathcal{W}$ due to $\Delta\state{\bm y}$ can be derived using Fr\'echet derivative as:
\begin{align}
    \Delta\mathcal{W} =& \int_{\domain}\state{\mrm{w}}\base[\bs\xi'] \bm P_{\bs\xi'} : \Delta\bm F_{\bs\xi'}\dV[\bs\xi'] \nonumber\\
    =& \int_{\domain}\state{\mrm{w}}\base[\bs\xi'] P_{\bs\xi', iJ}\left[ \state\omega\base[\bs\xi]\delta_{il}\xi_Q K^{-1}_{\bs\xi', QR} \state{A}_{RJ}\base[\bs\xi'] + \delta[\bs\xi-\bs\xi'] \nabla_{y_l} \state{B}_{iJ}\base[\bs\xi']\right]\bullet\Delta \state{y}_l\dV[\bs\xi']\nonumber\\
    =& \int_{\domain}\int_{\domain}\state{\mrm{w}}\base[\bs\xi'] P_{\bs\xi', iJ} \left[\state\omega\delta_{il}\xi_Q K^{-1}_{\bs\xi', QR}\state{A}_{RJ} + \delta[\bs\xi-\bs\xi'] \nabla_{y_l} \state{B}_{iJ}\right]\cdot\Delta\state{y}_l \dV[\bs\xi] \dV[\bs\xi']\nonumber\\
    =& \int_{\domain}\left\{\left[\int_{\domain} \state{\omega}\state{\mrm{w}}P_{\bs\xi', lJ}\state{A}_{RJ}K^{-1}_{\bs\xi', QR} \dV[\bs\xi']\right]\xi_Q\right\}\cdot\Delta\state{y}_l\dV[\bs\xi] +\nonumber\\
    &\int_{\domain}\left[\int_{\domain}\state{\mrm{w}} \delta[\bs\xi-\bs\xi'] P_{\bs\xi', iJ} \nabla_{y_l} \state{B}_{iJ} \dV[\bs\xi']\right] \cdot \Delta\state{y}_l\dV[\bs\xi] \nonumber\\
    =& \left\{\left[\int_{\domain} \state{\omega}\base\state{\mrm{w}}\base[\bs\xi']\bm P_{\bs\xi'}\state{\bm A}^T\base[\bs\xi'] \bm K_{\bs\xi'}^{-1}\dV[\bs\xi']\right]\bs\xi\right\}\bullet \Delta\state{\bm y}+\nonumber\\ &\left[\int_{\domain}\state{\mrm{w}}\base[\bs\xi'] \delta[\bs\xi-\bs\xi'] \bm P_{\bs\xi'}: \nabla_{\bm y}\state{\bm B}\base[\bs\xi'] \dV[\bs\xi']\right] \bullet\Delta\state{\bm y}.
    \label{eq:w_T_y}
\end{align}

According to the work conjugate relation, the unified force density state in the unified bond-associated correspondence formulation is obtained as
\begin{tcolorbox}
\begin{align}
    \state{\bm T}\langle\bs\xi\rangle = \left[\int_{\domain} \state{\omega}\base[\bs\xi]\state{\mrm{w}}\base[\bs\xi']\bm P_{\bs\xi'}\state{\bm A}^T\base[\bs\xi']\bm K_{\bs\xi'}^{-1}\dV[\bs\xi']\right]\bs\xi + \int_{\domain}\state{\mrm{w}}\base[\bs\xi'] \delta[\bs\xi-\bs\xi'] \bm P_{\bs\xi'}:\nabla_{\state{{\bm y}}}\state{\bm B}\base[\bs\xi'] \dV[\bs\xi'].
    \label{eq:def_T}
\end{align}
\end{tcolorbox}
The force density state presented in Eq.~\eqref{eq:def_T} is in a general form in terms of $\state{\bm A}\base[\bs\xi']$ and $\state{\bm B}\base[\bs\xi']$ without any assumptions of their specific forms. In the following part of this section, the specific expressions corresponding to different bond-associated deformation gradients are presented.

\begin{itemize}
    \item \textbf{Conventional model}
\end{itemize}

Considering the conditions presented in Eq.~\eqref{eq:Cond_Conv}, the force density state for the conventional model has the following form as
\begin{align}
    \state{\bm T}\base = \left[\int_{\domain} \state{\omega}\base[\bs\xi]\state{\mrm{w}}\base[\bs\xi']\bm P_{\bs\xi'}\base[\bs\xi']\bm K_{\bs\xi'}^{-1}\dV[\bs\xi']\right]\bs\xi
    = \state\omega\bm P\bm K^{-1}\bs\xi.
    \label{eq:conv_force}
\end{align}

Since the deformation gradient tensors and shape tensors are identical for each bond within the same horizon, the PK1 stress tensors are also identical. In this sense, the weighted average of $\bm P_{\bs\xi'}\bm K_{\bs\xi'}^{-1}$ is identical to the ones associated to every bond. Therefore, it can be simplified as $\bm P\bm K^{-1}$.

\begin{itemize}
    \item \textbf{Sub-horizon-based models}
\end{itemize}

For sub-horizon-based models subject to the conditions specified in Eq.~\eqref{eq:Cond_Sub}, the evaluation of the force density state is also straightforward. Instead of using the full horizon, the force density state is integrated or weight-averaged in a sub-horizon as
\begin{align}
    \state{\bm T}\langle\bs\xi\rangle =\left[\int_{h_{\bs\xi}}\state\omega\base\state{\mrm{w}}\base[\bs\xi']\bm P_{\bs\xi'} \bm K_{\bs\xi'}^{-1}\dV[\bs\xi']\right] \bs\xi.
    \label{eq:def_T_sub}
\end{align}

It is obvious that the sub-horizon plays an important role in computing both the deformation gradient tensors and the force density states associated with bonds. Hence, the choice of sub-domain is crucial.

In the work of Chowdhury et al.~\cite{chowdhury2019modified}, the whole horizon $\domain$ is divided into several non-overlapping sub-horizons $h_{\bs\xi'}$. The bond-associated deformation gradient $\bm F_{\bs\xi'}$ and shape tensor $\bm K_{\bs\xi'}$, hence the PK1 stress $\bm P_{\bs\xi'}$, are calculated for each sub-horizon and assumed to be identical for all the bonds within the same sub-horizon. Therefore, for this case, the force density state becomes
\begin{align}
    \state{\bm T}\langle\bs\xi\rangle = \state\omega\base \left[\int_{h_{\bs\xi}}\state{\mrm{w}}\base[\bs\xi']\dV[\bs\xi']\right] \bm P \bm K^{-1}\bs\xi.
\end{align}

Assuming the weight function takes the form of $\state{\mrm{w}}\base[\bs\xi']=1/V_{\domain}$, where $V_{\domain}$ is the volume of the total horizon, the above force density state expression can be further simplified as
\begin{align}
     \state{\bm T}\langle\bs\xi\rangle = \frac{V_{h_{\bs\xi'}}}{V_{\domain}} \state\omega\base \bm P \bm K^{-1}\bs\xi,
     \label{eq:force_sub1}
\end{align}
where $V_{h_{\bs\xi'}}$ represents the volume of sub-horizon.

In the initial work by Chen et al.~\cite{chen2018bond,chen2019stability,chan2021wave}, the sub-horizon for a bond overlaps with the sub-horizons of neighboring bonds within the same horizon. Assuming that the strain energy of each sub-horizon only depends on the target bond, Chen et al. derived the same force density state as given in Eq.~\eqref{eq:force_sub1}. Later, Chen and Chan~\cite{chan2022reformulation} reformulated the force density state by removing the above assumption and derived the general form for the bond force density state presented in Eq.~\eqref{eq:def_T_sub}.

\begin{itemize}
    \item \textbf{Projection-based model}
\end{itemize}

As for the the conditions stated in Eq.~\eqref{eq:Cond_Proj} for projection-tensor-based model, the expression for the bond force density state involves the gradient of non-zero state $\state{\bm B}$. Carrying out the calculation, the bond force density state can be derived as
\begin{align}
    \state{\bm T}\langle\bs\xi\rangle = \state\omega\base \left[\int_{\domain}\state{\mrm{w}}\base[\bs\xi'] \bm P_{\bs\xi'} \left(\unity[2] - \frac{\bs\xi'\otimes\bs\xi'}{|\bs\xi'|^2}\right) \bm K_{\bs\xi'}^{-1} \dV[\bs\xi'] \right] \bs\xi + \frac{\state{\mrm{w}}\base}{|\bs\xi|^2}\bm P_{\bs\xi}\bs\xi.
\end{align}

\begin{itemize}
    \item \textbf{Lagrangian-multiplier-based model}
\end{itemize}

Based on the updated form for the bond-associated deformation gradient given in Eq.~\eqref{eq:def_F_Lag2} for the Lagrangian-multiplier-based model, the bond force density state can be obtained from the general form as
\begin{align}
    \state{\bm T}\langle\bs\xi\rangle = \left[ \int_{\domain}\state\omega(\bs\xi, \bs\xi')\base \state{\mrm{w}}\base[\bs\xi']\bm P_{\bs\xi'}\bm K_{\bs\xi'}^{-1}\dV[\bs\xi']\right]\bs\xi
    \label{eq:force_Lag}
\end{align}

When $\lambda=1$ in Eq.~\eqref{eq:new_weight}, this expression will degenerate to the one for the conventional model (Eq.~\eqref{eq:conv_force}).

\begin{itemize}
    \item \textbf{Non-spherical-influence-function-based model}
\end{itemize}

The bond force density state for the non-spherical-influence-function-based model shares the same form as that of the Lagrangian-multiplier-based model, except the influence functions are different. The non-spherical influence function $\state\omega(\bs\xi, \bs\xi')\base$ for this model is given in Eq.~\eqref{eq:non-spherical}.

\section{Momentum Balance and Objectivity}
The equation of motion for the unified framework of bond-associated material correspondence models is the same as that for the conventional model proposed by Silling et al.~\cite{silling2007state}. The equation of motion is expressed as
\begin{tcolorbox}
    \begin{align}
        \rho(\bm X)\Ddot{\bm u}(\bm X, t) = \int_{\domain}\bigg\{\state{\bm T} \left[\bm X, t\right]\base - \state{\bm T}\left[\bm X', t\right]\base[-\bs\xi]\bigg\}\dV + \bm b(\bm X, t),
        \label{eq:pd_eom}
    \end{align}
\end{tcolorbox}
\noindent
where $\rho(\bm X)$ is the mass density of material point $\bm X$; $\Ddot{\bm u}(\bm X, t) := \partial^2\bm u / \partial t^2$ represents the second order derivative of displacements $\bm u$ with respect to time $t$; and $\bm b(\bm X, t)$ is the body force density of material point $\bm X$.

The above equation of motion (Eq.~\eqref{eq:pd_eom}) can be derived from the perspective of energy. Given an arbitrary domain $\mathcal{B}$, the total kinetic energy $K(t)$, the total external work $U(t)$ and the total strain energy $\Phi(t)$ can be computed by
\begin{align}
    K(t) = \frac{1}{2}\int_{\mathcal{B}}\rho\dot{\bm u}\cdot\dot{\bm u}\dV[\bm X],\quad U(t) = \int_0^t \int_{\mathcal{B}} \bm b \cdot \dot{\bm u}\dV[\bm X]\,\mrm{d}t,\quad \Phi(t) = \int_{\mathcal{B}}\mathcal{W}\dV[\bm X],
\end{align}
where $\dot{\bm u}:=\partial u/\partial t$ is the first order derivative of displacements with respect to time $t$.

As a result of conservation of energy, and assuming $K(0)=U(0)=\Phi(0)=0$ when $t=0$, the following two equations must be held,
\begin{align}
    U(t) =& \Phi(t) + K(t),\quad \forall t\geq 0\\
    \dot{U}(t) =& \dot\Phi(t) + \dot{K}(t),\quad \forall t\geq 0\label{eq:dU=dPhi+dK}
\end{align}
where
\begin{align}
    \dot{U}(t) =& \int_{\mathcal{B}}\bm b\cdot\dot{\bm u}\dV[\bm X] \label{eq:dotU}\\
    \dot{K}(t) =& \frac{1}{2}\int_{\mathcal{B}}\left[\rho\Ddot{\bm u}\cdot\dot{\bm u}+\rho\dot{\bm u}\cdot\Ddot{\bm u}\right]\dV[\bm X] = \int_{\mathcal{B}} \rho\Ddot{\bm u}\cdot\dot{\bm u}\dV[\bm X]\quad\label{eq:dotK}\\
    \dot{\Phi}(t) =& \int_{\mathcal{B}}\dot{\mathcal{W}}\dV[\bm X]
\end{align}
denote the time derivative of $U(t), K(t)$ and $\Phi(t)$, respectively. The $\dot{\Phi}(t)$ can be further expanded in terms of the bond force density state $\state{\bm T}\base$ and bond deformed state $\state{\bm y}\base$ as 
\begin{align}
    \dot{\Phi}(t) =& \int_{\mathcal{B}}\dot{\mathcal{W}}\dV[\bm X] = \int_{\mathcal{B}}\state{\bm T}\base \bullet \dot{\state{\bm y}}\base\dV[\bm X]\nonumber\\
    =&\int_{\mathcal{B}}\int_{\mathcal{B}}\state{\bm T}\base\cdot \dot{\state{\bm y}}\dV\dV[\bm X]\nonumber\\
    =&\int_{\mathcal{B}}\int_{\mathcal{B}}\state{\bm T}\base[\bm X' - \bm X] \cdot(\dot{\bm u}' - \dot{\bm u})\dV[\bm X']\dV[\bm X]\nonumber\\
    =&\int_{\mathcal{B}}\int_{\mathcal{B}}\state{\bm T}\base[\bm X' - \bm X] \cdot\dot{\bm u}'\dV[\bm X]\dV[\bm X'] - \int_{\mathcal{B}}\int_{\mathcal{B}}\state{\bm T}\base[\bm X' - \bm X] \cdot\dot{\bm u}\dV[\bm X]\dV[\bm X']\nonumber\\
    =& \int_{\mathcal{B}}\int_{\mathcal{B}}\Big[\state{\bm T}\base[\bm X' - \bm X] - \state{\bm T}'\base[\bm X - \bm X']\Big]\cdot\dot{\bm u} \dV[\bm X'] \dV[\bm X]\nonumber\\
    =& \int_{\mathcal{B}}\bigg\{\int_{\domain}\Big[\state{\bm T}\base - \state{\bm T}'\base[-\bs\xi]\Big]\dV\bigg\}\cdot\dot{\bm u}\dV[\bm X].
    \label{eq:dotPhi}
\end{align}
Substituting Eqs.~\eqref{eq:dotU}, \eqref{eq:dotK} and \eqref{eq:dotPhi} back into Eq.~\eqref{eq:dU=dPhi+dK} and applying the localization theorem, the equation of motion shown in Eq.~\eqref{eq:pd_eom} can be obtained.

\subsection{Balance of Linear Momentum}
For a bounded body $\mathcal{B}$ subjected to a body force density field $b(\bm X, t)$, the balance of linear momentum is always held. The proof is provided below.
\begin{align}
    &\int_{\mathcal{B}}\rho(\bm X)\Ddot{\bm u}(\bm X, t) - \bm b(\bm X, t)\dV[\bm X]\nonumber\\
    =& \int_{\mathcal{B}}\int_{\domain}\bigg\{\state{\bm T} \left[\bm X, t\right]\base - \state{\bm T}\left[\bm X', t\right]\base[-\bs\xi]\bigg\}\dV\dV[\bm X]\nonumber\\
    =&  \int_{\mathcal{B}}\int_{\mathcal{B}}\bigg\{\state{\bm T} \left[\bm X, t\right]\base - \state{\bm T}\left[\bm X', t\right]\base[-\bs\xi] \bigg\}\dV[\bm X']\dV[\bm X]\nonumber\\
    =& \int_{\mathcal{B}}\int_{\mathcal{B}}\state{\bm T} \left[\bm X, t\right]\base \dV[\bm X']\dV[\bm X] - \int_{\mathcal{B}}\int_{\mathcal{B}}\state{\bm T}\left[\bm X', t\right]\base[-\bs\xi] \dV[\bm X']\dV[\bm X]\nonumber\\
    =& \int_{\mathcal{B}}\int_{\mathcal{B}}\state{\bm T} \left[\bm X, t\right]\base \dV[\bm X']\dV[\bm X] - \int_{\mathcal{B}}\int_{\mathcal{B}}\state{\bm T}\left[\bm X, t\right]\base[\bs\xi] \dV[\bm X']\dV[\bm X]\nonumber\\
    =& 0
\end{align}
Since $\state{\bm T}\base = \bm 0$ whenever $\bs\xi\notin\domain$, the limit of integration on the inner integral may be changed from $\domain$ to $\mathcal{B}$. In the meantime, the order of integration can be exchanged. Therefore, the linear momentum balance is satisfied.

\subsection{Balance of Angular Momentum}
The angular momentum balance is another important law of physics that must be followed. For a bounded body $\mathcal{B}$, the angular momentum writes
\begin{align}
    \int_{\mathcal{B}}\state{\bm y}\base \times \left[\rho(\bm X)\Ddot{\bm u}(\bm X, t) - \bm b(\bm X, t) \right]\dV[\bm X] = \int_{\mathcal{B}}\state{\bm y}\base \times \state{\bm T}\base\dV.
\end{align}
Similarly, the force density state $\state{\bm T}\base=\bm 0$ if the bond $\bs\xi$ falls outside the horizon $\domain$. Thus, the integration domain can be changed from $\mathcal{B}$ to $\domain$. After substituting Eq.~\eqref{eq:def_T} into $\state{\bm T}\base$, the angular momentum yields 
\begin{align}
    &\left(\int_{\domain}\state{\bm y}\base \times \state{\bm T}\base\dV\right)_i \nonumber\\
    =&e_{ijl}\int_{\domain}\state{y}_j\left[\state{\omega}\int_{\domain} \state{\mrm{w}}\base[\bs\xi']P_{\bs\xi', lQ}\state{A}_{QR}^T \base[\bs\xi'] K_{\bs\xi', RS}^{-1}\dV[\bs\xi']\xi_S\right]\dV + \nonumber\\
    &e_{ijl}\int_{\domain}\state{y}_j\left[ \int_{\domain}\state{\mrm{w}}\base[\bs\xi'] \delta[\bs\xi-\bs\xi'] P_{\bs\xi', mQ}\state{ B}_{mQ,l}\base[\bs\xi'] \dV[\bs\xi']\right]\dV\nonumber\\
    =&e_{ijl}\int_{\domain} \state{\mrm{w}}\base[\bs\xi'] \left[\int_{\domain} \state{\omega} \state{y}_j\xi_S \dV[\bs\xi]\right]K_{\bs\xi',SR}^{-1} P_{\bs\xi', lQ} \state{A}_{QR}^T\base[\bs\xi'] \dV[\bs\xi'] + \nonumber\\
    &e_{ijl}\int_{\domain}\state{y}_j\left[ \int_{\domain}\state{\mrm{w}}\base[\bs\xi']\delta[\bs\xi-\bs\xi'] P_{\bs\xi', mQ} \state{B}_{mQ,l}\base[\bs\xi'] \dV[\bs\xi']\right]\dV\nonumber\\
    =&e_{ijl}\int_{\domain} \state{\mrm{w}}\base[\bs\xi'] F_{\bs\xi', jR}P_{\bs\xi', lQ}\state{A}_{QR}^T\base[\bs\xi'] \dV[\bs\xi'] + \nonumber\\
    &e_{ijl}\int_{\domain}\state{y}_j\left[ \int_{\domain}\state{\mrm{w}}\base[\bs\xi'] \delta[\bs\xi-\bs\xi'] P_{\bs\xi', mQ}\state{ B}_{mQ,l}\base[\bs\xi'] \dV[\bs\xi']\right]\dV.
    \label{eq:amb_general}
\end{align}
Eq.~\eqref{eq:amb_general} is a generalized expression for the angular momentum based on the unified deformation gradient and bond force density state. This expression will be examined for different cases of the bond-associated models.

\begin{itemize}
    \item \textbf{$\state{\bm A}\base = \unity[2]$} \quad and \quad \textbf{$\state{\bm B}\base = \bm 0$}
\end{itemize}

$\state{\bm A}\base = \unity[2]$ and $\state{\bm B}\base = \bm 0$ is the most common conditions in the bond-associated correspondence models, including the conventional model (Eq.~\eqref{eq:Cond_Conv}), the sub-horizon-based model (Eq.~\eqref{eq:Cond_Sub}), the Lagrangian-multiplier-based model (Eq.~\eqref{eq:Cond_Lag2}) and the non-spherical-influence-function-based model (Eq.~\eqref{eq:Cond_Nonsph}). For these models, Eq.~\eqref{eq:amb_general} can be simplified as 
\begin{align}
    \left(\int_{\domain}\state{\bm y}\base \times \state{\bm T}\base\dV\right)_i
    =& e_{ijl}\int_{\domain} \state{\mrm{w}}\base[\bs\xi'] F_{\bs\xi', jR}P_{\bs\xi', lQ}\delta_{QR} \dV[\bs\xi']\nonumber\\
    =& e_{ijl}\int_{\domain} \state{\mrm{w}}\base[\bs\xi'] F_{\bs\xi', jR} P_{\bs\xi', lR}\dV[\bs\xi']\nonumber\\
    =& e_{ijl}\int_{\domain} \state{\mrm{w}}\base[\bs\xi]F_{\bs\xi', jR} S_{\bs\xi', RQ}F^T_{\bs\xi', Ql}\dV[\bs\xi']\nonumber\\
    =& 0.
\end{align}
Therefore, the angular momentum balance is obtained for bond-associated models when $\state{\bm A}\base = \unity[2]$ and $\state{\bm B}\base = \bm 0$.

\begin{itemize}
    \item \textbf{$\state{\bm A}\base \neq \unity[2]$} \quad and \quad \textbf{$\state{\bm B}\base \neq \bm 0$}
\end{itemize}

For bond-associated models that $\state{\bm A}\base \neq \unity[2]$ and $\state{\bm B}\base \neq \bm 0$ in general, such as the projection-based model (Eq.~\eqref{eq:Cond_Proj}), the angular momentum balance can be derived in a slightly different manner. For the case of projection-based model, plugging the conditions given in Eq.~\eqref{eq:Cond_Proj} into Eq.~\eqref{eq:amb_general}, the angular momentum becomes
\begin{align}
     &\left(\int_{\domain}\state{\bm y}\base \times \state{\bm T}\base\dV\right)_i \nonumber\\
     =&e_{ijl}\int_{\domain} \state{\mrm{w}}\base[\bs\xi'] F_{\bs\xi', jR}P_{\bs\xi', lQ}\left(\delta_{QR} - \frac{\xi'_Q \xi'_R}{|\bs\xi'|^2}\right) \dV[\bs\xi'] + \nonumber\\
    &e_{ijl}\int_{\domain}\state{y}_j\left[ \int_{\domain}\frac{\state{\mrm{w}}\base[\bs\xi']}{|\bs\xi'|^2} \delta[\bs\xi-\bs\xi'] P_{\bs\xi', lQ}\xi'_Q \dV[\bs\xi']\right]\dV\nonumber\\
    =& e_{ijl}\int_{\domain} \state{\mrm{w}}\base[\bs\xi'] F_{\bs\xi', jR}P_{\bs\xi', lQ}\delta_{QR} \dV[\bs\xi'] - e_{ijl}\int_{\domain} \state{\mrm{w}}\base[\bs\xi'] F_{\bs\xi', jR}P_{\bs\xi', lQ} \frac{\xi'_Q \xi'_R}{|\bs\xi'|^2} \dV[\bs\xi'] + \nonumber\\
    &e_{ijl}\int_{\domain}\state{y}_j\frac{\state{\mrm{w}}}{|\bs\xi|^2}  P_{\bs\xi, lQ}\xi_Q \dV\nonumber\\
    =& e_{ijl}\int_{\domain} \state{\mrm{w}}\base[\bs\xi'] F_{\bs\xi', jR}P_{\bs\xi', lQ}\delta_{QR} \dV[\bs\xi'] - e_{ijl}\int_{\domain} \frac{\state{\mrm{w}}\base[\bs\xi']}{|\bs\xi'|^2} \state{y}_j \base[\bs\xi'] P_{\bs\xi', lQ}\xi'_Q\dV[\bs\xi'] + \nonumber\\
    &e_{ijl}\int_{\domain}\frac{\state{\mrm{w}}}{|\bs\xi|^2}\state{y}_j  P_{\bs\xi, lQ}\xi_Q \dV\nonumber\\
    =& e_{ijl}\int_{\domain} \state{\mrm{w}}\base[\bs\xi'] F_{\bs\xi', jR}P_{\bs\xi', lQ}\delta_{QR} \dV[\bs\xi']\nonumber\\
    =& 0.
\end{align}
Therefore, the angular momentum balance is also obtained for bond-associated models when \textbf{$\state{\bm A}\base \neq \unity[2]$} and \textbf{$\state{\bm B}\base \neq \bm 0$}.

\subsection{Objectivity}
To show the objectivity of the bond-associated correspondence models, let us assume there is a rigid body rotation $\bm Q$ superposed on the domain $\mathcal{B}$. Accordingly, the bond-associated deformation gradient for the body becomes
\begin{align}
    \bm F^+_{\bs\xi'} = \bm Q\bm F_{\bs\xi'}.
\end{align}

From the invariance of strain energy density under rigid body motion, it can be deduced that the PK1 stress after imposing rigid body rotation, denoted as $\bm P^+_{\bs\xi'}$, should be $\bm Q\bm P_{\bs\xi'}$, i.e.,
\begin{align}
    \Psi(\bm F_{\bs\xi'}) = \Psi(\bm Q\bm F_{\bs\xi'}) \quad \rightarrow \quad\bm P_{\bs\xi'}: \bm F_{\bs\xi'} \equiv \bm P^+_{\bs\xi'}: \bm F^+_{\bs\xi'}\quad \rightarrow \quad \bm P^+_{\bs\xi'} = \bm Q\bm P_{\bs\xi'}.
\end{align}

Replacing $\bm P_{\bs\xi'}$ with $\bm P^+_{\bs\xi'}$ in the unified force density state shown in Eq.~\eqref{eq:def_T} yields
\begin{align}
    \state{\bm T}^+\base =& \state{\omega}\base[\bs\xi]\left[\int_{\domain} \state{\mrm{w}}\base[\bs\xi']\bm P^+_{\bs\xi'}\state{\bm A}^T\base[\bs\xi']\bm K_{\bs\xi'}^{-1}\dV[\bs\xi']\right]\bs\xi + \int_{\domain}\state{\mrm{w}}\base[\bs\xi'] \delta[\bs\xi-\bs\xi'] \bm P^+_{\bs\xi'}:\nabla_{\bm y}\state{\bm B}\base[\bs\xi'] \dV[\bs\xi']\nonumber\\
    =& \state{\omega}\base[\bs\xi]\left[\int_{\domain} \state{\mrm{w}}\base[\bs\xi']\bm Q\bm P_{\bs\xi'}\state{\bm A}^T\base[\bs\xi']\bm K_{\bs\xi'}^{-1}\dV[\bs\xi']\right]\bs\xi + \int_{\domain}\state{\mrm{w}}\base[\bs\xi'] \delta[\bs\xi-\bs\xi'] \bm Q\bm P_{\bs\xi'}:\nabla_{\bm y}\state{\bm B}\base[\bs\xi'] \dV[\bs\xi'] \nonumber\\
    =& \bm Q\left\{\state{\omega}\base[\bs\xi]\left[\int_{\domain} \state{\mrm{w}}\base[\bs\xi']\bm P_{\bs\xi'}\state{\bm A}^T\base[\bs\xi']\bm K_{\bs\xi'}^{-1}\dV[\bs\xi']\right]\bs\xi + \int_{\domain}\state{\mrm{w}}\base[\bs\xi'] \delta[\bs\xi-\bs\xi'] \bm P_{\bs\xi'}:\nabla_{\bm y}\state{\bm B}\base[\bs\xi'] \dV[\bs\xi']\right\}\nonumber\\
    =& \bm Q \state{\bm T}\base.
\end{align}

Apparently, the transformation from state $\state{\bm T}\base$ to state $\state{\bm T}^+\base$, corresponding to the transformation from $\bm F_{\bs\xi'}$ to $\bm F^+_{\bs\xi'}$, follows the standard rules of tensor analysis. Therefore, the bond force density state $\state{\bm T}\base$ is objective.

\section{Summary}
This paper presents a unified framework of the bond-associated peridynamic material correspondence models. The main contributions of this work are summarized as follows:
\begin{enumerate}
    \item The bond-associated models were proposed to improve the accuracy of the nonlocal deformation gradient in mapping bond deformation between two distinct deformation configurations. As a result, the map between bond-associated deformation gradient and bond deformation states becomes injective. In this study, a unified model for the bond-associated deformation gradients was developed by generalization using two new state variables \textbf{$\state{\bm A}\base$} and \textbf{$\state{\bm B}\base$}. By choosing different values for these two state variables, different bond-associated models were recovered.
    \item Assuming strain energy equivalence with the conventional continuum mechanics theory, the unified force density state was derived using the Fr\'echet derivative based on the unified bond-associated deformation gradient. By choosing different values for the two state variables \textbf{$\state{\bm A}\base$} and \textbf{$\state{\bm B}\base$}, the bond-associated force density states for corresponding bond-associated deformation gradients were recovered.
    \item A systematic proof of balance of linear and angular momentum and objectivity of the unified framework was conducted in this study. This provides more theoretical support for those bond-associated correspondence models from physical perspectives.
\end{enumerate}

The unified framework developed in this study sheds light on how bond-associated correspondence models are inherently connected. It offers possibilities for the development or invention of new bond-associated correspondence models.

\newpage
\bibliographystyle{unsrt}
\bibliography{ref}
\end{document}